\documentclass[prd,twocolumn,showpacs,amsmath]{revtex4}
\usepackage{amssymb,amsmath}
\usepackage{graphicx}
\usepackage{color}

\begin{document}
\title{Cosmology in $(1+1)$-dimensional Ho\v rava-Lifshitz theory of gravity}

\author{J. P. M. Pitelli}
\email[]{joao.pitelli@ufabc.edu.br}
\affiliation{Centro de Matem\'atica, Computa\c{c}\~ao e Cogni\c{c}\~ao, UFABC, 09210-70, Santo Andr\'e, SP, Brazil}
\begin{abstract}

The $(1+1)$-dimensional Friedmann-Robertson-Walker (FRW) universe filled with a perfect fluid with equation of state $p=\omega \rho$ is analyzed through the view of Ho\v rava-Lifshitz (HL) theory of gravity. In this theory, the anisotropic scaling of space and time breaks Lorentz invariance of General Relativity (GR) in such a way that the gravitational action is no longer a topological invariant and the theory becomes dynamical. With the introduction of a perfect fluid through Schutz formalism, it is shown that the resulting dynamical theory  is very similar to the two-dimensional Jackiw-Teitelboim (JT) model, where a dilatonic degree of freedom is introduced to force a dynamical theory.  However, in HL theory, the introduction of a dilaton field is not necessary.

\end{abstract}

\maketitle

\section {Introduction}

Recently, Ho\v rava proposed a new theory of gravity \cite{horava1} based on an anisotropic scaling of the spatial and time coordinates,  which leads to the breaking of the relativistic invariance at short distances, with GR being recovered for low-energy limits. The main ingredient of this theory is  the anomalous scaling relation
\begin{equation}
t\to b^zt, \,\,\, x\to bx^{i},\,\,\,(i=1,2,\dots,d)
\label{anisotropic scaling}
\end{equation}
where $x^{i}$ and $t$ are the Minkowskian space and time coordinates, $b$ is a scale parameter and $z$ denotes the dynamical critical exponent, which should be $\geq d$ for the theory to be power counting renormalizable \cite{visser1} ($z=1$ corresponds to the relativistic scaling). Relation (\ref{anisotropic scaling}) clearly singles out a time coordinate, breaking the spacetime diffeomorphism invariance. However, this Lorentz Symmetry (LS) breaking poses no problem if the theory does not predict violation at energy scales which have already been tested \cite{visser1,sotiriou}.

To deal with the violation of LS, we foliate the spacetime, with  each leaf being a space hypersurface with constant $t$ and impose diffeomorphism invariance on such hypersurfaces. In this way, the framework for HL gravity is the Arnowitt-Deser-Misner (ADM) formalism \cite{adm}, in which the metric is  splitted in terms of the d-dimensional metric $h_{ij} (t, x)$ of the spatial slices of constant t, the lapse function $N(t, x)$, and the shift vector $N_i(t, x)$. One important assumption on the lapse function is called ``projectability condition'', namely $N(t,x)\equiv N(t)$, which simplifies considerably the HL gravitational action but generally gives rise to a unique integrated Hamiltonian. Nevertless, although it is not a necessary condition (there are non-projectable theories \cite{Blas}), it poses no problem in the study of FRW spacetimes, since the Hamiltonian in this case will be truly local \cite{Bellorin,Sotiriou2}. In this paper we will work with the projectability condition.

When applied to cosmology, HL theory showed some interesting consequences. Due to the presence of higher order spatial curvature terms, nonsingular bouncing universes were found \cite{Wang2,Calcagni,Brandenberger} and it was argued that HL theory may represent an alternative to inflation, since it might solve the flatness and horizon problems and generate scale-invariant perturbations either with or without inflation \cite{Wang3,Kiritsis,Mukohyama}. Also, when quantized, the higher order spatial curvature terms in HL quantum cosmology introduces a repulsive potential which solves the arbitrariness in the choice of boundary conditions at the big-bang singularity. In this way, HL quantum cosmology is uniquely solvable given an initial condition for the universe \cite{pitelli}.

Two-dimensional models of gravity have attracted a lot of attention in the last two decades. It is known that spherical reduction of four dimensional Einstein gravity leads to an effective two-dimensional theory with dilaton fields,  as well as string inspired theories \cite{Teitelboim,Cangemi}. Since quantization of gravity in four dimensions faces enormous difficulties, its natural to turn to simpler models by the lowering of the number of dimensions in models which share some features of the four-dimensional one. But before going to the quantum theory of gravity in two-dimensions, it is natural to consider the classical aspects of such theory. However, Einsten's gravity in two-dimensional spacetime is nondynamical, since the Einstein-Hilbert action is a topological invariant in the sense that its corresponding Lagrangean is a total derivative. In this way, in order to have a dynamical theory of gravity in a two-dimensional spacetime, an extra degree of freedom must be inserted in the gravitational action. Such model in which the analog of the four-dimensional vacuum Einsteins equations with a cosmological constant, namely $R-\Lambda$, is obtained from a local principle and the action is invariant is JT model \cite{Henneaux} given by the action
\begin{equation}
S=\int{d^2x\eta\left(R-\Lambda\right)},
\end{equation}
where $\eta$ is a scalar field (dilaton).
%It is usual to consider a dilaton, such as in the JT model \cite{Henneaux}.

The introduction of a dilaton in HL theory seems to be unnecessary since the action fails to be a topological invariant, giving rise to dynamical equations of motion \cite{wang}. In this way, it is possible to have a dynamical theory without the introduction of an extra degree of freedom. In this context we will analyze the FRW universe filled with a perfect fluid which couples to gravity via Schutz formalism \cite{schutz1,schutz2}. We will find solutions for the scale factor depending on the equation of state $p=\omega \rho$, where $\omega=0$ corresponds to dust and $\omega=1$ corresponds to a radiation fluid. Then we will compare the encountered solutions with solutions obtained in the JT  model \cite{cadoni}. We will see that, with a simple redefinition of constants, we obtain the same evolution of the scale factor on both models.

This paper is organized as follows: In Sec. II, the HL action in $(1+1)$-dimensions is derived. In Sec. III I couple the fluid to gravity via Schutz formalism and derive the equations of motion of the resulting action. Solutions of the equations of motion derived in Sec. III are presented and discussed in Sec. IV for a few cases. In Sec. V, the comparison between HL and JT theories is made and final discussions  are presented in Sec. V.

\section{HL action in $(1+1)$-dimensional FRW universe}

In $(1+1)$-dimensions, the gravitational action in HL theory is given by (here we follow the same notation as Ref. \cite{wang})
\begin{equation}
S_{HL}=\int{dtdx N(t)\sqrt{h_{11}}\left(\mathcal{L}_K-\mathcal{L}_V\right)},
\label{action_horava}
\end{equation}
where $N(t)$ is the lapse function, $h_{11}$ is the spatial part of the metric defined on the leaves $t=\text{constant}$, $\mathcal{L}_K$ is the kinetic part of the action given by
\begin{equation}
\mathcal{L}_K=K^{11}K_{11}-\lambda K^2,
\label{kinetic part}
\end{equation}
where $\lambda$ is a dimensionless constant and $K_{11}$ denotes the extrinsic curvature of the spatial leaves (note that $\mathcal{L}_K$ is the most general term involving the extrinsic curvature tensor which is invariant under the group of diffeomorphisms of the spatial slices). The potential part of the action $\mathcal{L}_V=\mathcal{L}_V\left(^{(1)}R,\nabla_{1},a_{1}\right)$  depends on the cosmological constant, on  the curvature $^{(1)}R$ of the leaves, which vanishes identically in one dimension and on the quantity $a_{i}=\partial_i\ln{N}$, which transform as vectors on the leaves.

In Eq. (\ref{kinetic part}), the extrinsic curvature $K_{11}$  is given by
\begin{equation}
K_{11}=\frac{1}{2N}(-\dot{h}_{11}+2\nabla_{1}N_1),
\end{equation}
where dot denotes time derivative, $\nabla_{1}$ denotes the covariant curvature on the leaves $t=\text{constant}$ and $N_{1}$ is the shift vector. In order to be power counting renormalizable $z$ should be greater or equal to the spatial dimension (here $d=1$). Taking the minimal requirement $z=1$ we have \cite{Bellorin2}
\begin{equation}
\mathcal{L}_V=-2\Lambda-\alpha a_{i}a^{i},
\label{potential part}
\end{equation}
where $\alpha$ is a coupling constant.

By considering the homogeneous and isotropic FRW metric
\begin{equation}
ds^2=-N(t)^2dt^2+a(t)^2dx^2,
\label{FRW metric}
\end{equation}
we have
\begin{equation}\begin{aligned}
&K_{11}=-\frac{a\dot{a}}{N}\\
&K=h^{11}K_{11}=-\frac{\dot{a}}{Na}\\
&K^{11}K_{11}=K^2=\frac{\dot{a}^2}{N^2a^2},
\label{K's}
\end{aligned}\end{equation}
and the HL action reads
\begin{equation}\begin{aligned}
S_{HL}&=\int{dtdxNa\left[(1-\lambda)K^2+2\Lambda\right]}\\
&=\int{dtdx\left[(1-\lambda)\frac{\dot{a}^2}{Na}+2\Lambda a N\right]}.
\label{action HL}
\end{aligned}\end{equation}
Note that the break of invariance determined by the constant $\lambda$ leads to a dynamical action. Contrary to GR, the kinetic part of the Lagrangean is not a total derivative.

\section{$(1+1)$-dimensional FRW universe filled with a perfect fluid}

In Schutz formalism \cite{schutz1,schutz2} for a perfect fluid in GR, the two-velocity of the fluid is expressed as \cite{Ahmed}
\begin{equation}
U_{\nu}=\mu^{-1}(\phi_{,\nu}+\theta S_{,\nu}),
\label{2velocity}
\end{equation}
where $\mu$ and $S$ are, respectively, the specific enthalpy and the specific entropy of the fluid. The potentials 	$\phi$
and $\theta$ have no clear physical meaning. With the usual normalization $U^{\nu}U_{\nu}$, Schutz showed that the action for the fluid is given by
\begin{equation}
S_{f}=\int{d^2x\sqrt{-g}p},
\end{equation}
where $p$ is the fluid's pressure, which is related with the fluid's density by $p=\omega \rho$.

For the FRW metric (\ref{FRW metric}), the two-velocity of the fluid is given by $U_{\nu}=N\delta^{0}_{\nu}$, so that
\begin{equation}
\mu=\left(\frac{\dot{\phi}+\theta \dot{S}}{N}\right).
\label{mu}
\end{equation}

By thermodynamical considerations, Lapchinski and Rubakov \cite{Lapchinski} found that the pressure as a function of Schutz potentials is given by
\begin{equation}
p=\frac{\omega\mu^{1+1/\omega}}{(1+\omega)^{1+1/\omega}}e^{-S/\omega}.
\label{pressure}
\end{equation}

In this way, the total action for the FRW universe filled with a perfect fluid in $(1+1)$-dimensional HL theory is given by (we assume that matter couples to gravity in HL theory via Schutz's representation)
\begin{equation}\begin{aligned}
S=&\int{dtdx\left[(1-\lambda)\frac{\dot{a}^2}{Na}+2\Lambda aN\right.}\\&\left.+\frac{\omega}{(1+\omega)^{1+1/\omega}}\left(\frac{\dot{\phi}+\theta\dot{S}}{N}\right)^{1+1/\omega}Na e^{-S/\omega}\right].
\end{aligned}
\label{total action}
\end{equation}

Following the canonical formalism we have that the momenta conjugated to $a$, $\phi$ and $S$ are given by
\begin{equation}
\left\{\begin{aligned}
&p_a=2(1-\lambda)\frac{\dot{a}}{Na},\\
&p_\phi=\frac{a \mu^{1/\omega}}{(1+\omega)^{1/\omega}}e^{-S/\omega},\\
&p_{S}=\theta p_{\phi}.
\label{momenta}
\end{aligned}\right.
\end{equation}
The Hamiltonian is then given by
\begin{equation}
\begin{aligned}
H&=p_a\dot{a}+p_\phi\left(\dot{\phi}+\theta\dot{S}\right)-L,\\
&=N\left(\frac{ap_a^2}{4(1-\lambda)}+\frac{p_\phi^{1+\omega}e^{S}}{a^{\omega}}-2\Lambda a\right).
\end{aligned}
\label{hamiltonian1}
\end{equation}
With the following canonical transformation
\begin{equation}
\left\{\begin{aligned}
&T=-p_Se^{-S}\phi^{-(1+\omega)},\\
&p_T=p_{\phi}^{1+\omega}e^{S},\\
&\bar{\phi}=\phi+(1+\omega)\frac{p_S}{p_\phi},\\
&\bar{p}_\phi=p_\phi,
\label{canonical transformation}
\end{aligned}\right.
\end{equation}
we have
\begin{equation}
H=N\left(\frac{ap_a^2}{4(1-\lambda)}+\frac{p_T}{a^{\omega}}-2\Lambda a\right).
\label{hamiltonian2}
\end{equation}

Varying the above Hamiltonian with respect to $N$ gives the Hamiltonian constraint
\begin{equation}
\mathcal{H}=\left(\frac{ap_a^2}{4(1-\lambda)}+\frac{p_T}{a^{\omega}}-2\Lambda a\right)\approx 0.
\label{hamiltonian constraint}
\end{equation}
The equations of motion which follow from the Hamilton's equation are given by
\begin{equation}
\left\{\begin{aligned}
&\dot{p}_T=0\Rightarrow p_T=\text{constant},\\
&\dot{T}=\frac{N}{a^{\omega}},\\
&\dot{p}_{a}=N\left[-\frac{p_a^2}{4(1-\lambda)}+2\Lambda +\omega \frac{p_T}{a^{1+\omega}}\right],\\
&\dot{a}=N\frac{ap_a}{2(1-\lambda)}
\label{equations of motion}
\end{aligned}\right.
\end{equation}

From now on, let us work in the cosmic time gauge ($N(t)=1$). By using the Hamiltonian constraint (\ref{hamiltonian constraint}) we have
\begin{equation}
\begin{aligned}
&\dot{p}_a=\frac{(\omega+1)p_T}{a^{1+\omega}}, \\
&\dot{a}=\frac{ap_a}{2(1-\lambda)}.
\end{aligned}
\label{equations cosmic time}
\end{equation}
If we take the derivative of the second equation in (\ref{equations cosmic time}) and use the first, together with the Hamiltonian constraint (\ref{hamiltonian constraint}), we arrive at
\begin{equation}
\frac{\ddot{a}}{a}=\frac{2\Lambda}{1-\lambda}+\frac{(\omega-1)p_T}{2(1-\lambda) a^{1+\omega}}.
\label{equation of motion a}
\end{equation}

\section{Solutions of the scale factor}

In this section we consider two possible fluids given by $\omega=0$ (dust) and $\omega=1$ (radiation). For a complete analysis we invite the reader to see Ref. \cite{cadoni}.

The first case we consider here is a radiation fluid with equation of state $p=\rho$ ($\omega=1$). Then Eq. (\ref{equation of motion a}) is simply given by
\begin{equation}
\frac{\ddot{a}}{a}=\frac{2\Lambda}{1-\lambda},
\end{equation}
with solution
\begin{widetext}
\begin{equation}\begin{aligned}
&a(t)=A \cosh{\left(\sqrt{\frac{2\Lambda}{1-\lambda}}t\right)}+ B \sinh{\left(\sqrt{\frac{2\Lambda}{1-\lambda}}t\right)},\,\,\, \text{if}\,\,\,\frac{2\Lambda}{1-\lambda}>0,\\
&a(t)=A+B t\,\,\,\text{if},\,\,\,\Lambda=0,\\
&a(t)=A \cos{\left(\sqrt{\left|\frac{2\Lambda}{1-\lambda}\right|}t\right)}+ B \sin{\left(\sqrt{\left|\frac{2\Lambda}{1-\lambda}\right|}t\right)},\,\,\, \text{if}\,\,\,\frac{2\Lambda}{1-\lambda}<0,\\
\end{aligned}
\end{equation}
\end{widetext}
where $A$ and $B$ are constants. Note that in this case, geometry has been completely decoupled from matter and the evolution of the universe does not depend on any matter parameters. For $\frac{2\Lambda}{1-\lambda}>0$ the universe is locally de Sitter and  begins with a big bang singularity if $|A|<|B|$ and with a finite size otherwise. For $\Lambda=0$ the universes begins with a big bang singularity at $t_0=-A/B$ and the scale factor behaves as $a(t)\sim t$ as $t\to\infty$. For $\frac{2\Lambda}{1-\lambda}<0$ we have periodic solutions of universes which begins with a big-bang singularity and recollapses after a finite time.

Considering now a fluid of dust $p=0$ ($\omega=0$) filling the universe we have
\begin{widetext}
\begin{equation}\begin{aligned}
&a(t)=A \cosh{\left(\sqrt{\frac{2\Lambda}{1-\lambda}}t\right)}+ B \sinh{\left(\sqrt{\frac{2\Lambda}{1-\lambda}}t\right)}+\frac{p_T}{4\Lambda},\,\,\, \text{if}\,\,\,\frac{2\Lambda}{1-\lambda}>0,\\
&a(t)=A+B t-\frac{p_T}{2}t^2\,\,\,\text{if},\,\,\,\Lambda=0,\\
&a(t)=A \cos{\left(\sqrt{\left|\frac{2\Lambda}{1-\lambda}\right|}t\right)}+ B \sin{\left(\sqrt{\left|\frac{2\Lambda}{1-\lambda}\right|}t\right)}+\frac{p_T}{4\Lambda},\,\,\, \text{if}\,\,\,\frac{2\Lambda}{1-\lambda}<0.\\
\end{aligned}
\end{equation}
\end{widetext}
For $\frac{2\Lambda}{1-\lambda}>0$, the universe is locally de Siter and can start with a big bang singularity or with a finite size depending on the parameters $A$, $B$, $p_T$ and $\Lambda$. For $\Lambda=0$, the universe recollapses after a finite time if $p_T>0$, but expands forever as $a(t)\sim t^2$ if $p_T<0$. For $\frac{2\Lambda}{1-\lambda}<0$ the universes oscillates and can be nonsingular if $p_T/(4\Lambda)>\sqrt{A^2+B^2}$.

\section{Comparison with Jackiw-Teitelboim model}

In Eq. (\ref{equation of motion a}), $p_T=p_\phi^{1+\omega}e^{S}$ by Eq. (\ref{canonical transformation}). By Eq. (\ref{momenta}) we have
\begin{equation}
\begin{aligned}
p_\phi^{1+\omega}e^{S}&=\left(\frac{a^{1+\omega}\mu^{1+\omega}}{(1+\omega)^{1+1/\omega}}e^{-S}e^{-S/\omega}\right)e^{S}\\
&=a^{1+\omega}p/\omega\equiv a^{1+\omega}\rho,
\end{aligned}
\label{density}
\end{equation}
where $\rho$ is the density of the fluid and we have used Eq. (\ref{pressure}).

Since the curvature for the metric (\ref{FRW metric}) is given by $R=2\frac{\ddot{a}}{a}$, Eq. (\ref{equation of motion a}) reads
\begin{equation}
R=\tilde{\Lambda}+\tilde{\rho},
\label{tilde}
\end{equation}
with the definitions $\tilde{\Lambda}\equiv\frac{4\Lambda}{(1-\lambda)}$ and $\tilde{\rho}\equiv \frac{\omega-1}{(1-\lambda)}\rho$.

In Ref. \cite{cadoni}, Cadoni and Mignemi studied the cosmology of the two-dimensional JT model through the action
\begin{equation}
S_{JT}=\int{d^2x}\sqrt{-g}\left[\eta \left(R-\Lambda\right)+\mathcal{L}_M\right],
\end{equation}
where $\mathcal{L}_M$ is the Lagrangean for the matter content. In analogy with the four-dimensional case, they take $\mathcal{L}_M=-\rho$ for the minimally coupled matter and $\mathcal{L}_M=-\eta \rho$ for the conformally coupled case.

Varying the action with relation to the dilaton field and the metric gives
\begin{itemize}
\item[a)] Minimally coupled matter:
\begin{equation}
\begin{aligned}
&R=\Lambda,\\
&-(-\nabla_\mu\nabla_\nu-g_{\mu\nu}\nabla^2)\eta+\frac{\Lambda}{2}g_{\mu\nu}\eta=\frac{1}{2} T_{\mu\nu}.
\end{aligned}
\label{minimally}
\end{equation}
\item[b)] Conformally coupled matter:
\begin{equation}
\begin{aligned}
&R=\Lambda+ \rho,\\
&-(-\nabla_\mu\nabla_\nu-g_{\mu\nu}\nabla^2)\eta+\frac{\Lambda}{2}g_{\mu\nu}\eta=\frac{1}{2} \eta T_{\mu\nu},
\end{aligned}
\label{conformally}
\end{equation}
\end{itemize}
where $T_{\mu\nu}=pg{\mu\nu}+(\rho+p)u_{\mu}u_\nu$.

Conservation of the energy-momentum tensor implies that $\dot{\rho}=-(p+\rho)\frac{\dot{a}}{a}$. With the equation of state $p=\omega \rho$ we have $\rho=M/a^{1+\omega}$, where $M$ is an integration constant. In this way, for the conformally coupled case, the equation for the scale factor becomes
\begin{equation}
R=\Lambda+\frac{M}{a^{1+\omega}}
\label{equation comparison}
\end{equation}

Note that  if we take $\omega<1$  in Eq. (\ref{tilde}) we get the same evolution for the scale factor as in Eq. (\ref{equation comparison}) for the conformally coupled case, but with the the substitutions $\Lambda\to\frac{4\Lambda}{(1-\lambda)}$ and $M\to\frac{\omega-1}{(1-\lambda)}p_T$. For $\omega=1$ we get the same evolution as in Eq. (\ref{minimally}) for the minimally coupled case. Note, however, that the fluid's density in this case is important only to solve the dilaton degree of freedom, which is not present in HL cosmology.

%In Ref. \cite{cadoni}, finite size universe emerge thanks to the presence of the dilaton field. In HL model we do not see such universes.
\acknowledgements
The author acknowledge stimulating discussions with A. Saa. This work was partially supported by FAPESP grant 2013/09357-9.

\section{Discussions}

The break of LS in HL theory of gravity in two-dimensions leads to  a dynamical action, contrary to what we see in GR theory, where the action is a topological invariant. With the introduction of a perfect fluid in the FRW universe through Schutz formalism and following a Hamiltonian prescription, the equation of motion for the scale factor is equivalent to the case of a conformally coupled dilatonic theory in JT model for $\omega<1$ and to the case of a  minimally coupled dilaton for $\omega=1$ with a simple redefinitions of parameters, where $p=\omega \rho$. However, since the dilatonic degree of freedom is unnecessary in HL model, the only dynamical quantity  is the scale factor.  The presence of a dilatonic degree of freedom in JT predicts more complex universes. In such model, the initial singularity  can be either of metric or dilatonic nature. In HL cosmological models the universes are simpler, since the scale factor is the unique dynamical quantity.

%In fact, in the JT model the universe can be singular because the scalar curvature or the dilaton field are singular. In HL theory we do not need to worry about such complex universes since a dilaton field seems to be unnecessary for a dynamical universe.
%It can be seen in Ref. \cite{cadoni} that the presence of a dilaton gives rise to universes with a finite size at its beginning, as well as universes which recollapses after a finite time. However, in HL theory there is only one equation of motion, namely the one for the scale factor [see Eq. (\ref{equation of motion a})].

\end{document}